\documentclass[conference]{IEEEtran}
\IEEEoverridecommandlockouts

\usepackage{cite}
\usepackage[hidelinks]{hyperref}
\usepackage{amsmath,amssymb,amsfonts}
\usepackage{algorithmic}
\usepackage{graphicx}
\usepackage{textcomp}
\usepackage{xcolor}
\def\BibTeX{{\rm B\kern-.05em{\sc i\kern-.025em b}\kern-.08em
    T\kern-.1667em\lower.7ex\hbox{E}\kern-.125emX}}

\usepackage{booktabs}
\usepackage{makecell}
\usepackage{subcaption}

\begin{document}
\bstctlcite{IEEEtran:BSTcontrol}

\title{CREDIT: Cost-guided Reduction-reuse with Efficient DSMEM Inter-CTA Tiling}

\author{
\IEEEauthorblockN{Zhengxiong Li}
\IEEEauthorblockA{\textit{University of Wisconsin--Madison} \\
Madison, WI, USA \\
zhengxiong.li@wisc.edu}
\and
\IEEEauthorblockN{Tsung-Wei Huang}
\IEEEauthorblockA{\textit{University of Wisconsin--Madison} \\
Madison, WI, USA \\
tsung-wei.huang@wisc.edu}
\and
\IEEEauthorblockN{Umit Ogras}
\IEEEauthorblockA{\textit{University of Wisconsin--Madison} \\
Madison, WI, USA \\
uogras@wisc.edu}
}

\maketitle

\begin{abstract}
NVIDIA distributed shared memory (DSMEM) enables direct shared-memory access within a thread block cluster. However, cluster synchronization, remote access, and resource costs make it difficult to determine when DSMEM improves performance. To fill this gap, we propose CREDIT, a cost-guided framework that identifies DSMEM-profitable workload patterns, predicts their profitability range, and delivers consistent speedups across diverse workloads. 
CREDIT combines three innovations: 
(1) a profiling-driven characterization that identifies workload patterns likely to benefit from DSMEM; (2) a transformation that applies DSMEM to reduction-reuse workloads; (3) a cost model based on profiling data, to determine its profitability range. Evaluations on diverse workloads show CREDIT achieves 91.7\% prediction accuracy on profitability. CREDIT beats \texttt{torch.compile}, Triton, and optimized non-DSMEM CUDA baselines on all six workloads, with geometric-mean speedups of $1.466\times$ on RTX~5090 and $1.318\times$ on H100. CREDIT's source code is publicly available at \texttt{https://github.com/zhengxiongli08/CREDIT}.
\end{abstract}

\section{Introduction}
\label{sec:intro}
Modern GPUs increasingly expose the memory wall as peak performance has grown faster than off-chip memory bandwidth~\cite{Gholami2024}. Table~\ref{tab:gpu-balance} summarizes this trend under the Roofline model~\cite{Williams2009} for representative NVIDIA GPUs, including Volta~\cite{nvidia_v100_datasheet}, Ampere~\cite{nvidia_a100}, Hopper~\cite{nvidia_h100, nvidia_hgx}, and Blackwell~\cite{nvidia_dgx_b200,nvidia_rtx_blackwell}. The dense FP16 compute-to-bandwidth ratio rises from 139 FLOPs/byte on V100 to about 295 FLOPs/byte on H100, and Blackwell GPUs remain in the same regime. As a result, many operators fail to achieve peak performance unless they can reduce memory footprint to increase arithmetic intensity. 

Starting from Hopper, NVIDIA introduced thread block clusters and DSMEM~\cite{nvidia_hopper,nvidia_cuda_dsmem}. A cluster is a group of cooperative thread arrays (CTAs) that are co-scheduled on streaming multiprocessors (SMs) and can synchronize as a unit. DSMEM allows these blocks to directly load, store, and perform atomic operations on one another's shared memory (SMEM). However, DSMEM operations are not free. Remote shared-memory accesses are slower than local shared memory accesses, cluster synchronization adds latency, and larger shared memory allocations or fixed cluster sizes can reduce occupancy and performance.

\begin{table}[t]
    \centering
    \caption{Compute-to-memory scaling in representative NVIDIA GPUs}
    \label{tab:gpu-balance}
    \begin{tabular}{llccc}
        \toprule
        \textbf{GPU} & \textbf{Arch.} &
        \makecell{\textbf{Peak Perf.}\\\textnormal{(TFLOP/s)}} &
        \makecell{\textbf{Mem. BW}\\\textnormal{(TB/s)}} &
        \makecell{\textbf{Roofline Ridge}\\\textnormal{(FLOPs/byte)}} \\
        \midrule
        V100 SXM    & Volta     & 125  & 0.900 & \textbf{139} \\
        A100 SXM    & Ampere    & 312  & 2.039 & \textbf{153} \\
        H100 SXM    & Hopper    & 990  & 3.350 & \textbf{295} \\
        B200 SXM    & Blackwell & 2250 & 8.000 & \textbf{281} \\
        RTX 5090    & Blackwell & 419  & 1.792 & \textbf{234} \\
        \bottomrule
    \end{tabular}
\end{table}

We propose CREDIT, which identifies which workload patterns benefit from DSMEM, predicts the profitability range for a given shape, and applies the transformation only when it pays off.
Supported by detailed profiling data on DSMEM performance, we identify that DSMEM pays off for wide reduction-reuse kernels, which reduce a wide vector and then reuse its per-element data to compute a full-size output. DSMEM can reduce off-chip memory traffic by exchanging data across a cluster, avoiding rereading from global memory between two reduction passes. CREDIT applies a transformation to split the input data across multiple blocks, where each block computes its own partial results and exchanges with others within the cluster. CREDIT also involves a cost model to determine whether it's beneficial to apply the transformation to a given workload and problem size. 

We evaluate CREDIT across diverse workloads. The results show size- and architecture-dependent crossovers: no-DSMEM baselines are usually preferable for small problem sizes, while CREDIT becomes competitive when the avoided global-memory rereads amortize extra overheads introduced by DSMEM, including synchronization, DSMEM communication, and occupancy costs. Evaluations show CREDIT outperforms the fastest baseline on all six workloads, with a geometric-mean speedup of $1.466\times$ on RTX~5090 and $1.318\times$ on H100. Nsight Compute measurements demonstrate CREDIT reduces off-chip memory traffic by 33-60\% on different workloads.

Our contributions in this paper are outlined as follows:
\begin{itemize}
    \item \textbf{Workload and DSMEM performance analysis}. We quantitatively characterize the performance of DSMEM in terms of read/write latency, throughput, and synchronization overhead, which indicates promising workload patterns.

    \item \textbf{Profitability cost model}. We develop a profile-guided cost model that compares the time we can save against extra overheads introduced by DSMEM, including cluster synchronization, DSMEM communication, and occupancy costs, to predict profitability. 

    \item \textbf{Cross-platform evaluation}. We evaluate diverse workloads on RTX~5090 and H100 against \texttt{torch.compile}, Triton, and optimized no-DSMEM baselines, and validate the off-chip traffic reduction using Nsight Compute.
\end{itemize}

For the rest, Section~\ref{sec:background} provides background and discusses prior works. Section~\ref{sec:methodology} presents the proposed CREDIT framework. Finally, Section~\ref{sec:evaluation} presents the evaluations on diverse workloads and Section~\ref{sec:conclusion} concludes the paper.

\section{Background and Related Work}
\label{sec:background}

\subsection{DSMEM Programming Model}
\begin{figure}[t]
    \centering

    \begin{subfigure}[t]{\linewidth}
        \centering
        \includegraphics[width=0.9\linewidth]{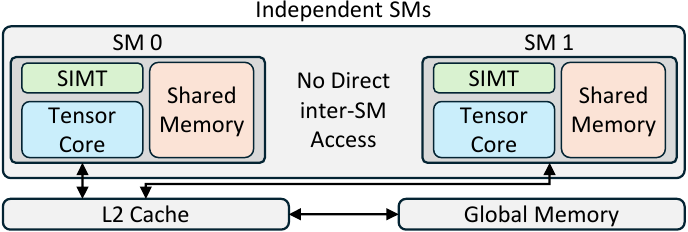}
        \caption{Conventional block-local access}
        \label{fig:no_dsmem}
    \end{subfigure}

    \begin{subfigure}[t]{\linewidth}
        \centering
        \includegraphics[width=0.9\linewidth]{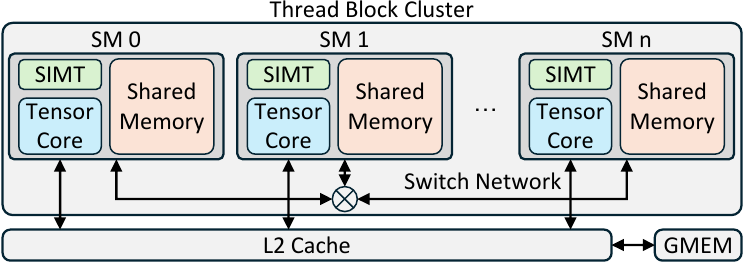}
        \caption{Cluster-scoped peer access}
        \label{fig:with_dsmem}
    \end{subfigure}

    \caption{Shared-memory access with and without a thread block cluster. DSMEM lets a CTA address allocations owned by peer CTAs through the inter-SM interconnect; each allocation remains physically local to its owner}
    \label{fig:comparison}
\end{figure}

In the conventional CUDA execution model, shared memory is scoped to one thread block, and blocks in a grid have no guaranteed relative scheduling order. As shown in Fig.~\ref{fig:no_dsmem}, communicating intermediate state across thread blocks generally requires device memory, atomics, or a kernel boundary. Starting from Hopper architecture, thread block clusters add an intermediate cooperation scope: all blocks in a cluster are scheduled concurrently on SMs within one graphics processing cluster (GPC) and can synchronize through cluster-wide barriers~\cite{nvidia_hopper,nvidia_cuda_dsmem}. Each block still owns a distinct shared-memory allocation, but DSMEM allows a block to map an address into a peer block's allocation and issue remote loads, stores, and atomic operations. The resulting address space is logically distributed across the cluster rather than a physically unified cache, as illustrated in Fig.~\ref{fig:with_dsmem}.

\subsection{Prior Approaches to Reducing Data Movement}
Off-chip traffic can first be reduced without cross-CTA communication. IO-aware algorithms restructure an operator around the memory hierarchy; FlashAttention~\cite{dao2022flashattention} and FlashAttention-2~\cite{dao2024flashattention} use tiled online reductions and refined work partitioning to avoid materializing the attention matrix in HBM. At the graph and compiler levels, TVM~\cite{tvm}, PyTorch~\cite{Ansel2024}, DNNFusion~\cite{Niu2021}, Ansor~\cite{ansor}, and Roller~\cite{roller} optimize fusion or schedules using generated code and performance models. AStitch~\cite{Zheng2022} exploits hierarchical reuse when fusing memory-intensive graphs, Welder~\cite{welder} uses a tile-traffic cost model to coordinate intra- and inter-operator reuse, and MonoNN~\cite{mononn} forms monolithic GPU kernels to reduce intermediate materialization and launch overhead. Triton~\cite{Tillet2019} provides a tiled programming model for custom kernels, while ThunderKittens~\cite{spector2025thunderkittens} exposes tile-level abstractions for registers, shared memory, asynchronous copies, and matrix units. These approaches make data movement a first-class optimization objective, but registers and per-CTA shared memory remain their principal scopes for on-chip communication and reuse.

Prior microarchitectural studies provide an empirical basis for reasoning about this additional scope. Luo et al.~\cite{Luo2024} and Luhnen et al.~\cite{Luehnen2024} benchmark Hopper's memory hierarchy and characterize DSMEM latency and bandwidth across cluster configurations, while Jin et al.~\cite{Jin2024} expose topology-dependent latency and bandwidth behavior in real GPU networks-on-chip. These studies establish that peak specifications are insufficient and inter-SM communication must be calibrated on physical hardware. CREDIT complements them by isolating the read, store, visibility, and barrier primitives exercised by its protocol on two architectures and feeding those measurements into workload selection.

Recent systems directly turn the cluster scope into workload optimizations. FlashFuser~\cite{Huang2026} introduces DSMEM communication abstractions, data movement analysis, and an analytical cost model to search fusion plans for compute-intensive operator chains. ClusterFusion~\cite{luo2025clusterfusion} profiles sensitivity to cluster configuration, models collective traffic, and uses optimized operators to fuse LLM decoding stages. Quack~\cite{quack} proposes a series of optimizations for memory-bound kernels on Hopper and Blackwell platforms, including RMSNorm, softmax, and cross-entropy loss.

\subsection{Positioning of CREDIT}
Prior work therefore spans conventional kernel fusion and dataflow optimization, hardware characterization, and target-specific DSMEM designs. CREDIT aims to answer whether an operator at a given shape should use clustered staging at all, relative to a competitive non-clustered implementation. We formalize this workload class and use matched primitive and control measurements together with one no-DSMEM timing to predict the profitability of each workload-shape pair and crossover. The contribution is a workload-level characterization and selection framework, validated on Hopper H100 and Blackwell RTX~5090 across both profitable and unprofitable cases, rather than the first use of DSMEM or a new cluster-reduction collective.

\begin{figure*}[t]
    \centering
    \includegraphics[width=\textwidth]{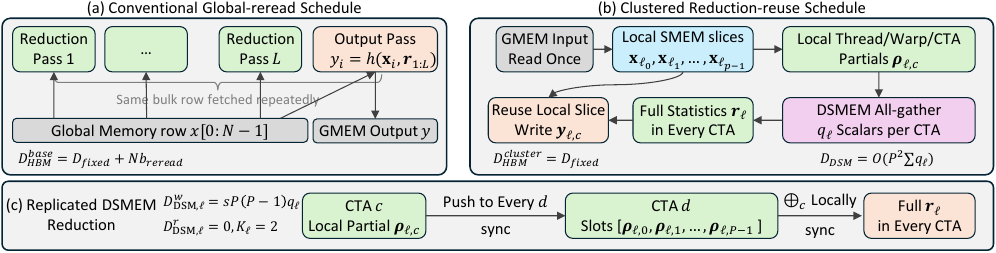}
    \caption{CREDIT transformation for reduction-reuse kernels. (a) Conventional passes reread bulk rows. (b) A cluster retains owner-local slices and exchanges scalar partials. (c) Each CTA pushes partials to every peer, which combines its local slots after synchronization}
    \label{fig:overview}
\end{figure*}

\section{CREDIT Framework}
\label{sec:methodology}

This section presents CREDIT using three key parts: (1) detailed profiling on the hardware costs introduced by DSMEM; (2) a transformation from multi-pass reduction into distributed local staging and compact reduction; (3) a cost model to determine the profitability for a given workload and problem size.

\subsection{DSMEM Primitive Characterization}
\label{sec:primitive-characterization}

We measure DSMEM performance with three CUDA microbenchmarks based on established GPU memory-hierarchy profiling methods~\cite{Mei2017,Jia2018}. The first makes every load wait for the preceding load to expose access latency. The second issues many independent reads or stores in parallel to measure their achieved rate. The third repeatedly synchronizes the cluster. Each microbenchmark forces its two CTAs onto different SMs. We report the median of 15 runs.

\begin{table}[h]
    \centering
    \caption{Profiling on RTX~5090 and H100}
    \label{tab:primitive-costs}
    \setlength{\tabcolsep}{2.5pt}
    \resizebox{0.98\columnwidth}{!}{%
    \begin{tabular}{lrrrrrrr}
        \toprule
        & \multicolumn{3}{c}{Latency (cycles)} & \multicolumn{4}{c}{Rate (byte/cycle)} \\
        \cmidrule(lr){2-4}\cmidrule(l){5-8}
        GPU & \makecell{Local\\load} & \makecell{Remote\\load} & Sync & \makecell{Local\\read} & \makecell{Remote\\read} & \makecell{Local\\store} & \makecell{Remote\\store} \\
        \midrule
        RTX~5090 & 34 & 216 & 404 & 20.89 & 3.84 & 33.01 & 2.14 \\
        H100 & 30 & 191 & 851 & 20.89 & 4.77 & 27.67 & 21.37 \\
        \bottomrule
    \end{tabular}
    }
\end{table}

Table~\ref{tab:primitive-costs} exposes two key facts. First, a remote load has about $6.4\times$ the latency and $4.4$--$5.4\times$ lower throughput than a local read. Therefore, only compact partial results should be moved between CTAs and read locally after synchronization; bulk data remains in its owner's SMEM. Second, synchronization and remote-store costs differ substantially between the two GPUs, so one fixed DSMEM penalty cannot represent both devices. The cost model in Section~\ref{sec:cost-model} uses the measured local-read, local-store, and remote-store rates, while a matched control kernel captures synchronization and scheduling overhead.

\subsection{Reduction-Reuse Transformation}
\label{sec:pattern}

\begin{table}[b]
    \centering
    \caption{Notation used throughout Section~\ref{sec:methodology}}
    \label{tab:methodology-notation}
    \scriptsize
    \begin{tabular}{@{}p{0.33\columnwidth}p{0.6\columnwidth}@{}}
        \toprule
        Symbol & Definition \\
        \midrule
        $M,N$ & Number of rows, and elements per row \\
        $P,L,q_\ell,s$ & Cluster CTAs; stages; stage-$\ell$ statistics; scalar bytes \\
        $b_{\mathrm{keep}},b_{\mathrm{reread}},b_B$ & Staged; eliminated-reread; baseline B/element \\
        $S_{\mathrm{CTA}},S_{\mathrm{scratch}},S_{\max}$ & Memory footprint; scratch; SMEM limit \\
        $S,f,W$ & SM count; device frequency; sequential cluster waves \\
        $\widehat B_{\mathrm{eff}},B_{\mathrm{SMEM}}^r,B_{\mathrm{DSM}}^w$ & Effective source; local-read; remote-store rates \\
        $D_{\mathrm{HBM}},D_{\mathrm{DSM}}^{w/r}$ & HBM and remote DSMEM write/read byte/row \\
        $T_B,T_{\mathrm{save}},T_{\mathrm{ctrl}}$ & Baseline; saved-reread; cluster-control times \\
        $T_{\mathrm{replay}},T_{\mathrm{deposit}},T_{\mathrm{DSM}}$ & Local-replay; deposit; remote-store times \\
        \bottomrule
    \end{tabular}
\end{table}

As shown in Fig.~\ref{fig:overview}(a), consider an input row of width $N$, with per-element inputs $\mathbf{x}_i$. At stage $\ell\in\{1,\ldots,L\}$, index $j\in\{1,\ldots,q_\ell\}$ selects one of the $q_\ell$ scalar statistics, $\mathbf{r}_{<\ell}$ denotes all statistics from earlier stages, $g_{\ell,j}$ is the per-element contribution, and $\bigoplus$ is an associative reduction such as sum or maximum. After all $L$ stages, the elementwise function $h$ produces the output:
\begin{equation}
    \begin{aligned}
    r_{\ell,j} &= \bigoplus_{i=0}^{N-1} g_{\ell,j}(\mathbf{x}_i,\mathbf{r}_{<\ell})
    & j&=1,\ldots,q_\ell \\
    y_i &= h(\mathbf{x}_i,\mathbf{r}_{1:L})
    & i&=0,\ldots,N-1
    \end{aligned}
    \label{eq:reduction-reuse}
\end{equation}

As shown in Fig.~\ref{fig:overview}(b), we partition each row into $P$ non-overlapping slices $\{\mathcal{I}_c\}_{c=0}^{P-1}$, one per CTA. CTA $c$ retains only its slice, and $\rho_{\ell,c,j}$ denotes its local partial for statistic $j$ at stage $\ell$:
\begin{equation}
    \rho_{\ell,c,j}=\bigoplus_{i\in\mathcal{I}_c}g_{\ell,j}(\mathbf{x}_i,\mathbf{r}_{<\ell}),
    \qquad r_{\ell,j}=\bigoplus_{c=0}^{P-1}\rho_{\ell,c,j}
    \label{eq:hierarchical-reduction}
\end{equation}

For retained bytes per element $b_{\mathrm{keep}}$, fixed per-CTA scratch $S_{\mathrm{scratch}}$, and the configured per-CTA dynamic-SMEM limit $S_{\max}$, staging is legal only if:
\begin{equation}
    S_{\mathrm{CTA}}(N,P)=\left\lceil\frac{N}{P}\right\rceil b_{\mathrm{keep}}+S_{\mathrm{scratch}}\leq S_{\max}
    \label{eq:capacity}
\end{equation}

Fig.~\ref{fig:overview}(a) and (b) compare this global-reread schedule with our clustered reduction-reuse schedule. 
Let $D_{\mathrm{fixed}}$ denote per-row HBM traffic that is identical in both schedules (e.g., output writes, which may themselves scale with $N$, but are not affected by clustering). If we denote the eliminated reread bytes per element by $b_{\mathrm{reread}}$, then:
\begin{equation}
    D_{\mathrm{HBM}}^{\mathrm{base}} = D_{\mathrm{fixed}} + Nb_{\mathrm{reread}},
    \qquad 
    D_{\mathrm{HBM}}^{\mathrm{cluster}} = D_{\mathrm{fixed}}
    \label{eq:hbm-traffic}
\end{equation}
For a memory-bound kernel, the traffic-only speedup ceiling is $D_{\mathrm{HBM}}^{\mathrm{base}}/D_{\mathrm{HBM}}^{\mathrm{cluster}}$. 

\subsection{Cluster Reduction and Reuse Protocol}
\label{sec:kernel-transformation}

At stage $\ell$, each CTA compresses values through thread accumulation, warp shuffles, and an SMEM reduction to obtain $\boldsymbol{\rho}_{\ell,c}=(\rho_{\ell, c, 1}, \ldots, \rho_{\ell, c, q_\ell})$ before performing any cluster-wide communication. Then, it pushes compact partials into peer-owned buffers so that the final combining reads are local. 

Fig.~\ref{fig:overview}(c) shows the replicated push/all-gather used by CREDIT. For source CTA $c$, destination CTA $d$, and statistic $j$, let $\mathrm{slot}_d[c,j]$ be the local-SMEM location in $d$ reserved for $c$, and let $r_{\ell,j}^{(d)}$ be the full statistic reconstructed by $d$. One of the first $P$ threads writes each partial into the corresponding remote slot:
\begin{equation}
    \mathrm{slot}_{d}[c,j]\leftarrow\rho_{\ell,c,j},\qquad
    r_{\ell,j}^{(d)}=\bigoplus_{c=0}^{P-1}\mathrm{slot}_{d}[c,j]
    \label{eq:replicated-reduction}
\end{equation}

The first barrier exposes pushed partials. Every CTA then combines only local slots; a second barrier protects scratch reuse. 
Unlike a design where a single root CTA gathers all partials and every peer then rereads the result, this replicated all-gather lets every CTA reconstruct the full statistic independently, avoiding both a serial bottleneck at the root and a second round of remote reads at the cost of replicated scalar stores. 
Let $D_{\mathrm{DSM}}^{w/r}$ denote total remote write and read bytes per row, let $s$ be the bytes per scalar statistic, and let $K$ count cluster-wide barriers per row. The protocol costs are
\begin{equation}
    D_{\mathrm{DSM}}^{w}=sP(P-1)\sum_{\ell=1}^{L}q_\ell,\quad D_{\mathrm{DSM}}^{r}=0,\quad K=2L
    \label{eq:dsmem-protocol-cost}
\end{equation}

The factor $P(P-1)$ counts $P$ source CTAs writing to their $P-1$ remote peers; subsequent reads are local because each destination owns its slots. CTA $c$ finally applies $h$ to the elements in its retained slice and writes its output segment.

\subsection{Profitability and Candidate Selection}
\label{sec:cost-model}

The model is evaluated for $M$ rows and each legal cluster size $P$. 
Using peak HBM or L2 bandwidth to predict how long the baseline takes to stream a row is unreliable since cache residency changes with row shape, and a one-CTA kernel achieves very different fractions of peak bandwidth on RTX~5090 versus H100.
Therefore, we use a single timing of the selected non-DSMEM CUDA path, $T_B(M,N)$. 
Since this timing requires only the baseline kernel, it can be measured before any clustered variant is generated or tuned, making it a cheap first screen rather than part of the search itself.
\begin{equation}
    \begin{aligned}
        \widehat{B}_{\mathrm{eff}}(M,N)&=\frac{MN b_B}{T_B(M,N)}\\
        T_{\mathrm{save}}(M,N)&=\frac{MN b_{\mathrm{reread}}}{\widehat{B}_{\mathrm{eff}}}=T_B(M,N)\frac{b_{\mathrm{reread}}}{b_B}
    \end{aligned}
    \label{eq:effective-bandwidth}
\end{equation}

Eq.~(\ref{eq:effective-bandwidth}) replaces a device peak with the rate actually achieved by the baseline at the same shape; it requires no DSMEM workload timing and has no fitted coefficient. 
We independently launch a work-free control kernel with the same grid, $P$, dynamic-SMEM footprint, and $K = 2L$ barriers. 
The excess time this control takes relative to a one-CTA control, $T_{\mathrm{ctrl}}(M,N,P)$, captures cluster launch, scheduling, occupancy, and synchronization costs.

When staging limits residency to one CTA per SM, a row cluster occupies $P$ SMs, so at most $\lfloor S/P\rfloor$ row clusters execute concurrently on a device with $S$ SMs. Let $W(M,P)$ be the resulting number of sequential cluster waves, $f$ the SM frequency, $B_{\mathrm{SMEM}}^r$ the achieved local-SMEM read rate, and $B_{\mathrm{DSM}}^w(P)$ the per-CTA remote-store issue rate for cluster size $P$. The incremental local replay and remote-store times are
\begin{equation}
    \begin{aligned}
        W(M,P) &= \left\lceil\frac{M}{\lfloor S/P\rfloor}\right\rceil\\
        T_{\mathrm{replay}} &= \frac{W(M,P)\lceil N/P\rceil b_{\mathrm{keep}}}{fB_{\mathrm{SMEM}}^r}\\
        T_{\mathrm{DSM}} &= \frac{sW(M,P)(P-1)\sum_{\ell}q_\ell}{fB_{\mathrm{DSM}}^w(P)}
    \end{aligned}
    \label{eq:model-costs}
\end{equation}

Because the $P$ source CTAs issue their stores concurrently on different SMs, their combined wall-clock cost is not $P$ times a single CTA's cost. 
So, although Eq.~(\ref{eq:dsmem-protocol-cost}) counts $P$ source CTAs for byte accounting, Eq.~(\ref{eq:model-costs}) does not multiply elapsed time by $P$.
$T_{\mathrm{replay}}$ charges the incremental local-SMEM read needed to apply $h$. 
The initial SMEM deposit is issued alongside the compulsory global load that streams each element from HBM. 
We call this load's duration $T_{\mathrm{input}}$, the source-stream time.
Let $T_{\mathrm{SMEM}}^w$ be the deposit's standalone local-store issue time and define $[z]^+=\max(z,0)$. We charge only the non-overlapped residual $T_{\mathrm{deposit}}=[T_{\mathrm{SMEM}}^w-T_{\mathrm{input}}]^+$, which is zero for both measured devices because local-SMEM store issue is faster than the per-SM input stream. The predicted benefit is
\begin{equation}
    \Delta T(M,N,P)\approx T_{\mathrm{save}}-T_{\mathrm{ctrl}}-T_{\mathrm{replay}}-T_{\mathrm{deposit}}-T_{\mathrm{DSM}}
    \label{eq:profitability}
\end{equation}

We define $N^\star$ as the smallest tested width for which Eq.~(\ref{eq:capacity}) holds and $\Delta T>0$ for at least one legal $P$; final kernel tuning may choose a different legal $P$ because this is a pair and crossover predictor rather than a cycle-exact model. The non-DSMEM timing makes the model profile-guided, but evaluating three legal $P$ values is inexpensive relative to compiling and benchmarking three clustered kernels. 
The same reasoning rejects three cases by construction: elementwise and scalar-output reductions when $b_{\mathrm{reread}}\approx0$, scans when ordered dependencies grow $K$, and selection or stencil kernels when remote state grows with $N$ rather than $\sum_\ell q_\ell$. 
In each case, increasing $N$ does not help. A reduction with no reread has nothing to amortize; growing $K$ adds synchronization faster than width can offset it, and remote state that scales with $N$ defeats the compact-partial assumption.

\section{Evaluation and Discussion}
\label{sec:evaluation}

\begin{figure*}[tb]
    \centering
    \includegraphics[width=0.98\textwidth]{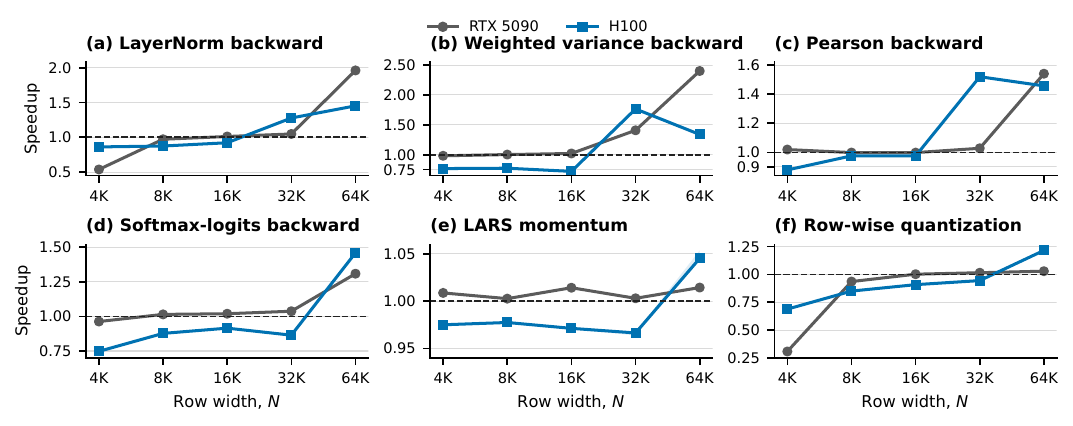}
    \caption{CREDIT's speedups over the fastest baseline across workloads and shapes}
    \label{fig:workload-scaling}
\end{figure*}

\subsection{Experimental Setup and Baselines}

We evaluate CREDIT on an NVIDIA RTX~5090 and an H100 SXM. Both use CUDA 13.0, PyTorch~2.11.0 and Triton~3.6.0. 
Our benchmark suite contains six reduction--reuse workloads with FP32 input format from five categories: LayerNorm~\cite{Ba2016} and weighted-variance backward, Pearson-correlation backward, softmax-logits backward, LARS momentum~\cite{You2017}, and row-wise int8 quantization. LayerNorm uses $M=2048$ rows; the others use $M=4096$. We sweep $N\in\{4096,8192,16384,32768,65536\}$ and legal cluster sizes $P\in\{2,4,8\}$, selecting the fastest measured $P$ for performance results.

We compare CREDIT against three non-DSMEM baselines: \texttt{torch.compile}~\cite{Ansel2024}, Triton~\cite{Tillet2019}, and a CUDA baseline that selects a local-SMEM-staged kernel when the row fits in one CTA, or a global-reread kernel otherwise.
The first two represent optimized framework and expert-DSL implementations, while the CUDA baseline provides an ablation mechanism. 
Reported times exclude \texttt{torch.compile} compilation time and Triton's autotuning search time; both are measured once and amortized outside the timed region. 
We report the speedup as $S_B=T_B/T_{\mathrm{CREDIT}}$, where $T_{\mathrm{CREDIT}}$ is the CREDIT kernel runtime. Thus, $S_B>1$ means CREDIT is faster than baseline $B$, and ``best'' denotes the fastest of the three baselines at that shape.

\subsection{Workload-by-Workload Scaling}

Fig.~\ref{fig:workload-scaling} shows width- and architecture-dependent crossovers. At 4K, CREDIT is slower than the one-CTA baseline (geometric-mean speedups are $0.738\times$ and $0.815\times$ for RTX~5090 and H100, respectively), because the one-CTA baseline exploits cache or local staging without cluster overhead.
At 64K, CREDIT beats the fastest baseline on all six workloads. Table~\ref{tab:baseline-summary} decomposes this result by baseline: the geometric-mean speedups over the fastest baseline are $1.466\times$ on RTX~5090 and $1.318\times$ on H100 across all workloads.

\begin{table}[t]
    \centering
    \caption{Geometric-mean CREDIT speedup at $N=64K$ over the baselines across all workloads}
    \label{tab:baseline-summary}
    \begin{tabular}{lcccc}
        \toprule
        GPU & \texttt{torch.compile} & Triton & CUDA & Best \\
        \midrule
        RTX~5090 & $1.617\times$ & $1.937\times$ & $1.980\times$ & $1.466\times$ \\
        H100 & $1.469\times$ & $1.525\times$ & $1.533\times$ & $1.318\times$ \\
        \bottomrule
    \end{tabular}
\end{table}

As expected, the crossover point changes as a function of the hardware and workload. RTX~5090 first exceeds parity at 16K and wins every workload at 32K, whereas H100 wins every workload only at 64K, demonstrating that device costs shift the amortization point. 
This result is consistent with Table~\ref{tab:primitive-costs}, where H100's higher barrier latency (851 vs.\ 404 cycles) raises $T_{\mathrm{ctrl}}$ and shifts the amortization point to wider rows.
Overall, CREDIT beats the fastest baseline at 22 points in Fig.~\ref{fig:workload-scaling} (i.e., workload-shape pairs) when running on RTX~5090, and 9 points when running on H100.

The workload spread reinforces this interpretation. Weighted variance provides the largest RTX~5090 gain, rising from $0.982\times$ at 4K to $2.401\times$ at 64K, while Pearson and softmax-logits exhibit later, moderate crossovers. 
LARS is the limiting positive case, reaching only $1.014\times$ and $1.045\times$ at 64K, and quantization crosses the best baseline at 16K on RTX~5090 but only at 64K on H100. Thus, width creates an opportunity, but reusable traffic, baseline efficiency, and cluster overhead determine whether CREDIT realizes it.

\subsection{Cost-Model Validation}

\begin{figure}[htbp]
    \centering
    \includegraphics[width=\columnwidth]{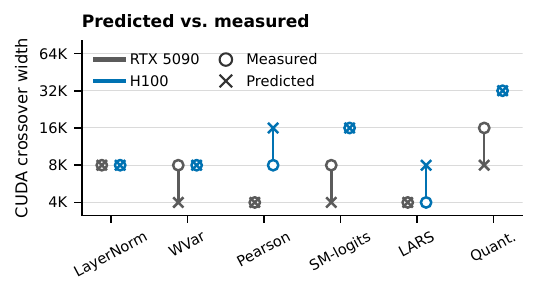}
    \caption{Cost-model validation. Crosses and open circles show the predicted and measured smallest width at which DSMEM beats matched non-DSMEM CUDA; vertical segments denote prediction error. Each vertical step doubles the row width}
    \label{fig:model-validation}
\end{figure}

Fig.~\ref{fig:workload-scaling} establishes practical performance against the fastest baseline, whereas the model asks the narrower mechanism question of whether CREDIT beats matched non-DSMEM CUDA without ever timing a DSMEM workload. 
We obtain local-SMEM and DSMEM rates from Table~\ref{tab:primitive-costs} in Section~\ref{sec:primitive-characterization}; the work-free control supplies $T_{\mathrm{ctrl}}$ for each $(M,N,P)$; the non-DSMEM kernel supplies $T_B$; and static analysis supplies $b_B$, $b_{\mathrm{keep}}$, $b_{\mathrm{reread}}$, and $\{q_\ell\}$.

\begin{table}[t]
    \centering
    \caption{Profitability-pair accuracy across six workloads and five shapes per GPU}
    \label{tab:model-ablation}
    \begin{tabular}{lccc}
        \toprule
        Model & RTX~5090 & H100 & Overall \\
        \midrule
        Peak bandwidth + serialized staging & 27/30 & 16/30 & 43/60 \\
        CREDIT: achieved bandwidth + overlap & 27/30 & 28/30 & 55/60 \\
        \bottomrule
    \end{tabular}
\end{table}

Table~\ref{tab:model-ablation} evaluates 60 profitability pairs: five widths, six workloads, and two GPUs. CREDIT reaches 90.0\% accuracy on RTX~5090, 93.3\% on H100, and 91.7\% overall. Replacing device peaks with the baseline's achieved bandwidth and overlapping the compulsory input load with the SMEM deposit improves H100 accuracy from 16/30 to 28/30 and overall accuracy from 43/60 to 55/60, without fitting a coefficient to DSMEM workload timings.
The five remaining errors occur only among the 32 points where the baseline instead uses the local-SMEM-staged kernel, at shapes near the staging boundary, with measured DSMEM/CUDA ratios of 0.996, 0.986, and 0.994 on RTX~5090 and 1.015 and 1.022 on H100, all within 2.2\% of parity. All 28 points using a pure global-reread CUDA baseline are classified correctly.

Fig.~\ref{fig:model-validation} provides a stricter crossover test. CREDIT's prediction is exact for seven of the 12 device--workload pairs and differs by one tested factor-of-two width for the other five; no crossover is missed by more than one interval. The result supports the intended use of Eq.~\ref{eq:profitability}: it is a mechanism screen that rejects clearly unprofitable shapes and narrows the tuning region, not a cycle-exact runtime predictor or a replacement for final cluster-size search. Identifying a mechanism opportunity does not by itself guarantee a win over a separately optimized framework schedule; practical deployment requires both the model screen and the fastest-baseline comparison in Fig.~\ref{fig:workload-scaling}.

Negative controls further validate the screening model's structural rules. One-pass reductions expose no recoverable reread traffic, scans introduce ordered dependencies, and selection or stencil kernels communicate noncompact state; none creates the owner-local reduction--reuse pattern required by CREDIT. Rejecting these cases before clustered code generation shows that CREDIT identifies opportunity from dataflow rather than vector width alone.

\subsection{Traffic Validation}

\begin{figure}[t]
    \centering
    \includegraphics[width=\columnwidth]{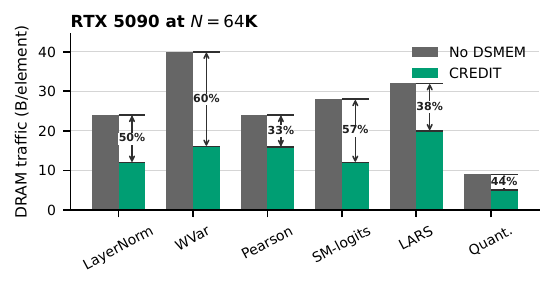}
    \caption{Nsight Compute DRAM traffic on RTX~5090 at $N=65536$. Bars compare matched non-DSMEM CUDA with CREDIT; labels give the traffic reduction from owner-local staging}
    \label{fig:traffic-validation}
\end{figure}

Fig.~\ref{fig:traffic-validation} independently validates the root cause of speedups enabled by CREDIT. CREDIT reduces measured RTX~5090 DRAM traffic by 33--60\%: LayerNorm drops from 24 to 12~B/element, weighted variance from 40 to 16, Pearson from 24 to 16, softmax-logits from 28 to 12, LARS from 32 to 20, and quantization from 9 to 5. Output traffic is unchanged; the reduction comes from retaining owner-local slices and replacing repeated global reads with compact statistics. 
Traffic savings alone do not predict the speedup: LARS removes 38\% of traffic but gains only 1--5\% over the best baseline at 64K, because the framework baseline is already substantially faster than matched CUDA. This is exactly why CREDIT values avoided source time at the baseline's achieved rate rather than at raw traffic reduction. It explicitly subtracts cluster control, replay, and DSMEM transport, while a purely traffic-based metric would overstate LARS's practical benefit.

\section{Conclusion}
\label{sec:conclusion}

DSMEM is a selective reduction-reuse mechanism, not a general fusion fabric. CREDIT correctly classifies 55/60 workload-shape pairs across RTX~5090 and H100. As problem size increases, CREDIT beats the best baseline on all six workloads, with geometric-mean speedups of $1.466\times$ and $1.318\times$, respectively. Clustered variants should be generated only when recovered source time amortizes distributed staging and compact inter-CTA reduction. Our future work will explore integrating CREDIT with more complex GPU workloads and graph models~\cite{SET, Taskflow}.

\vspace{1mm}
\noindent \textit{Disclosure}: Dr. Ogras is affiliated with Samsung Austin Research \& Development Center and Advanced Computing Lab (SARC/ACL). This relationship has been approved under applicable outside activities policies.

\bibliographystyle{IEEEtran}
\bibliography{references.bib}

\end{document}